\documentclass{article}
\usepackage{spconf,amsmath,graphicx,url,amsfonts}

\title{A Speech Representation Anonymization Framework via Selective Noise Perturbation}
\name{Minh Tran and Mohammad Soleymani}
\address{Institute for Creative Technologies, University of Southern California, Los Angeles, USA}
\begin{document}
%
\maketitle
\begin{abstract}
Privacy and security are major concerns when communicating speech signals to cloud services such as automatic speech recognition (ASR) and speech emotion recognition (SER). Existing solutions for speech anonymization mainly focus on voice conversion or voice modification to convert a raw utterance into another one with similar content but different, or no, identity-related information. However, an alternative approach to share speech data under the form of privacy-preserving representation has been largely under-explored. In this paper, we propose a speech anonymization framework that achieves privacy via noise perturbation to a selected subset of the high-utility representations extracted using a pre-trained speech encoder. The subset is chosen with a Transformer-based privacy-risk saliency estimator. We validate our framework on four tasks, namely, Automatic Speaker Verification (ASV), ASR, SER and Intent Classification (IC) for privacy and utility assessment. Experimental results show that our approach is able to achieve a competitive, or even better, utility compared to the speech anonymization baselines from the VoicePrivacy2022 Challenges, providing the same level of privacy. Moreover, the easily-controlled amount of perturbation allows our framework to have a flexible range of privacy-utility trade-offs without re-training any component.
\end{abstract}
\section{Introduction}
Collecting and sharing speech data from local devices are essential for cloud-based speech processing models. From the service providers' perspective, gaining access to more data enables them to train more robust models and provide better services. From the users' perspective, sharing personal data allows experiencing more personalized services such as healthcare \cite{chen2020fedhealth} or automatic speech recognition (ASR) \cite{mcgraw2016personalized}. However, privacy and security are major concerns when communicating personal speech data, as raw utterances often contain sensitive identifiable information such as ethnicity, gender, or even health condition. One promising line of solution focuses on client-side privacy \cite{wu2021understanding}, providing users full control to protect their privacy by sanitizing raw utterances locally before uploading the data for cloud-based processing without the need for a trusted server. 


Anonymization, which aims at removing identifiable information from speech signals while retaining other information, is a commonly used approach. The majority of speech anonymization methods focus on voice anonymization that either modifies audio inputs via signal processing techniques \cite{mcadams1984spectral, kai2021lightweight} or utilizes speech synthesis to perturb the x-vector \cite{snyder2018x} of the inputs \cite{srivastava2020design, wu2021understanding, prajapati2022voice}. On the other hand, existing studies on privacy-preserving representation learning for speech generally leverage adversarial training to remove identity-related information from representations, but are limited to a single target application such as ASR \cite{srivastava2019privacy} or emotion recognition \cite{jaiswal2020privacy}. To the best of our knowledge, there has been no prior work on privacy-preserving representation learning for multiple applications (as in this work).

In this study, we propose a speech anonymization framework to sanitize the representations extracted from pre-trained speech encoders. At the core of our approach is a Privacy-risk Saliency Estimator (PSE) that learns to predict the importance of individual representation positions from a speaker identification system. Based on the estimations from PSE, the top $k\%$ positions with the highest estimated privacy-risk would be perturbed by adding easily-controlled Laplacian noise. We validate our method on four tasks, namely, Automatic Speaker Verification (ASV) for privacy evaluation, Automatic Speaker Verification (ASR), Emotion Recognition (ER), and Intent Classification (IC) for utility performance. Our work provides preliminary evidence that using speech data in the form of extracted representations results in systems achieving competitive performance compared to the commonly-used voice conversion/modification approaches. Moreover, our approach enables a flexible privacy-utility trade-off.

\section{Proposed Method}
\begin{figure}[t]
  \centering
  \includegraphics[width=\linewidth]{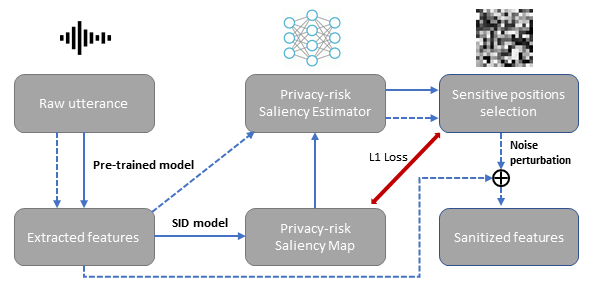}
  \caption{Overview of our framework. Solid lines show the workflow of the training process for the Privacy-risk Saliency Estimator. Dashed lines show the workflow during the inference process to produce a sanitized representation for a raw utterance.}
  \label{fig:overview}
\end{figure}

Figure \ref{fig:overview} shows a high-level overview of the proposed method. We first train a Speaker Identification model on a closed set of speakers with dataset $D$, which is used to construct a Saliency Map Dataset. We then train a Privacy-risk Saliency Estimator (PSE) to predict the privacy-salient positions given the extracted representations. The PSE module allows our framework to be extensible to unseen speakers outside $D$. Finally, we add controlled noise to selected positions within the input representations to produce sanitized output representations. 

\noindent \textbf{Privacy-risk Saliency Map Dataset.} The idea of saliency maps was originally proposed in computer vision to identify the most salient regions in images for trained classification models' decisions. Gradient-based methods construct the saliency map for a given image as the absolute value of the gradient of the input pixels with respect to the loss for the true class of the image, which is accessible via backpropagation. In this work, we use \textit{SmoothGrad} \cite{smilkov2017smoothgrad} to compute saliency maps. It is important to note that we can only compute saliency maps if ground-truth labels (speaker identities in this case) are available. We do not have this information during the inference process, which motivates the development of a model that can estimate the saliency maps directly from the input representations.

For a given 2D representation $x \in \mathbb{R}^{t \times d}$ of an utterance, we hypothesize that some positions in $x$ contain more sensitive information than the others (\textit{i.e.}, these positions can leak more identifiable information than the others). Note that these positions can change given different speech signals with different representations. We use \textit{SmoothGrad} to find these positions. Specifically, we first train a Speaker Identification (SID) model $M$ on a dataset $D=\{(x_i, y_i)\}$, with $x_i$ being the extracted features from a pre-trained model (\textit{e.g.}, \texttt{wav2vec2}) and $y_i$ being the identity of $x_i$. For each sample $(x_i, y_i)$ in $D$, we compute the privacy-risk saliency map $s_i$ with \textit{SmoothGrad}, which is the average of $\lvert \frac{\delta L_{y_i}}{\delta x_i + \epsilon} \rvert$, where $L_{y_i}$ is the loss of $M$ on the true class $y_i$ of $x_i$ with different versions of noise $\epsilon$. As a result, we construct a privacy-risk saliency map dataset $D'=\{(x_i,s_i)\}$ containing the same number of samples as $D$ with $x_i$ and $s_i$ of the same sizes. This dataset is then used to train our PSE module.\\
\noindent \textbf{Privacy-risk Saliency Estimator.} The purpose of the Privacy-risk Saliency Estimator (PSE) is to take in the extracted feature representations $x_i$ for an utterance $u$ and output an estimate for the privacy-risk saliency map $s'_i$ such that $s'_i \approx s_i$. A good PSE should be able to (a) model the dynamics of the saliency maps given different speech representations; and (b) generalize to unseen speakers outside $D$. We use a Transformer encoder \cite{vaswani2017attention} as our architecture for the PSE. As a high-level overview, the Transformer encoder is a stack of Transformer layers, each of which consists of a self-attention module followed by a feed-forward neural network. For a given $x_i \in \mathbb{R}^{t \times d}$, we expect the self-attention modules to capture information along the temporal dimension while the feed-forward neural networks capture important signals from the feature dimension, which enables the architecture to make precise position-level predictions. The model's parameters are optimized with the $L1$ Loss between the outputs of PSE and the ground-truth saliency maps extracted with a trained Speaker Identification model, as explained above. \\
\noindent \textbf{Noise Perturbation.} 
We assume that not all positions within the extracted representations from pre-trained encoders carry the same amount of speaker-related information. By injecting noise to the more \textit{sensitive} positions, we hope to reduce the sensitive information carried within these dimensions to achieve privacy. To improve the utility of the output representations, we keep the less sensitive, \textit{i.e., insensitive} positions, intact. In this study, we define the group of \textit{sensitive} positions as positions with the top $k\%$ highest privacy-risk values estimated by PSE. We later investigate the effect of $k$ on the privacy and utility of the sanitized outputs. Motivated by Differential Privacy and the \textit{Laplace mechanism} \cite{dwork2014algorithmic}, we set a fixed input-independent bound of $[-1,1]$ and add Laplacian noises $\stackrel{iid}{\sim} Lap(\frac{2}{\epsilon})$ independently to the selected positions.

\section{Experimental Setups}
In this study, we focus on sanitizing the extracted representations from \texttt{wav2vec2} \cite{baevski2020wav2vec}. However, the proposed framework should be applicable to other pre-trained speech encoders. Our implementations and training procedures follow the SUPERB benchmark \cite{yang2021superb} to enhance reproducibility. To show that the sanitized representations is applicable to multiple applications, we pick three downstream tasks with emphasis on different aspects of speech, namely, ASR for content, ER for paralinguistics, and IC for semantics. Our code is publicly available\footnote{\url{https://github.com/mtran14/dp_w2v2.git}}.

\subsection{Tasks, Datasets \& Models}
\textbf{VoxCeleb1 (SID \& ASV).} We use the VoxCeleb1 dataset \cite{nagrani2017voxceleb} for training the SID and ASV models. It contains approximately $150,000$ utterances from $1,211$ speakers collected from around $21,819$ videos on YouTube. For a fair privacy evaluation, we first split the dataset into two partitions. The first subset contains approximately $50,000$ utterances from $400$ speakers, and is used to train the SID model and construct the Privacy-risk Saliency Map dataset. The second subset, which contains around $100,000$ utterances from the remaining $811$ speakers, is used to train the ASV model for privacy evaluation. The output representations from \texttt{wav2vec2} is fed to  the SID model consisting of a mean-pooling layer followed by a fully-connected layer to make the speaker identity predictions. The SID model is trained using a cross-entropy loss. For ASV, we train the x-vector model \cite{snyder2018x} with an AMSoftmax loss \cite{wang2018additive}. Following prior work \cite{tomashenko2020introducing, wu2021understanding, kai2021lightweight}, we report the equal error rate (EER) as our privacy metric (a higher EER implies a better privacy-preserving representation).

\noindent \textbf{IEMOCAP (ER).} We use the IEMOCAP dataset \cite{busso2008iemocap} for training the ER model. The original dataset contains approximately $12$ hours of data from $10$ actors performing improvised or scripted scenarios that are designed to invoke emotions. Following the SUPERB benchmark's setting, we only keep four emotion classes (neutral, happy, sad, angry) from the IEMOCAP dataset to get a more balanced dataset. The ER model adds a mean-pooling layer followed by a fully-connected layer to the encoder for emotion recognition. We report the classification accuracy as the utility metric. 

\noindent \textbf{Fluent Speech Commands (IC).} We use the Fluent Speech Commands dataset \cite{lugosch2019speech} to train our IC model. The dataset contains around $30,000$ utterances from $97$ speakers with $31$ unique intents. Similar to ER, the IC model contains a mean-pooling layer followed by a fully-connected layer. We report the classification accuracy as the utility metric. 

\noindent \textbf{LibriSpeech (ASR).} We use the commonly used LibriSpeech audiobook dataset \cite{panayotov2015librispeech} to train our ASR model. Specifically, we use LibriSpeech's train-clean-100/dev-clean/test-clean subsets to train, validate and test our model. The training set contains more than $100$ hours of transcribed speech from $251$ speakers while both the validation and test set contains more than $5$ hours of transcribed speech from $40$ speakers. Our ASR model is a two-layer 1024-unit Bidirectional LSTM, trained with the CTC loss \cite{graves2006connectionist} on characters. We do not use any language model to improve the model's performance during inference. The Word Error Rate (WER) is used as the utility metric.
\vspace{-5mm}
\subsection{Baselines}
We use the baselines provided in the VoicePrivacy2022 Challenge \cite{tomashenko2022voiceprivacy}, in addition to a signal processing-based methods from \cite{kai2021lightweight} as our baseline. \\
\noindent \textbf{Baseline 1.} We use the first baseline provided in the VoicePrivacy2022 Challenge, which attempts to modify the x-vector for input speech signals. The method first extracts the x-vector \cite{snyder2018x}, the fundamental frequency (F0), and bottleneck features from the input speech signal. The method then modifies the extracted x-vector with an external pool of x-vectors before using a speech synthesizer to produce an anonymized speech from the extracted features. We use a variant of the first baseline with features extracted from a finetuned wav2vec2.0 model \cite{baevski2020wav2vec} and a HifiGAN-based speech synthesizer (more details can be found from Section 6.4.3 in \cite{tomashenko2022voiceprivacy}). \\
\noindent \textbf{Baseline 2.} We use the second baseline provided in the VoicePrivacy2022 Challenge, which is a signal processing approach using McAdams transformation \cite{mcadams1984spectral}. The McAdam Coefficient is uniformly sampled from a fixed range to get a randomized version of the anonymization method in \cite{patino2021speaker}.
\\
\noindent \textbf{Baseline 3.} We use the \textit{R+MS} method proposed by Kai \textit{et al.} \cite{kai2021lightweight}. It is a combination of Resampling and Modulation Spectrum Smoothing \cite{takamichi2015naist}, and is the method suggested by Kai \textit{et al.} to achieve strong privacy-utility trade-offs. 
\\
\noindent \textbf{Baseline 4.} We use the \textit{R+MS+M+CH+CL} method proposed by Kai \textit{et al.} \cite{kai2021lightweight}. The method consists of Resampling, Modulation Spectrum Smoothing, McAdams Transformation \cite{mcadams1984spectral}, clipping, and Chorus. It is the most secured approach from their provided toolkit \cite{kai2021lightweight}. 
\vspace{-5mm}
\subsection{Details.}
\noindent \textbf{Implementation.} Our PSE is a standard Transformer encoder, which consists of $6$ layers with a dropout rate of $0.1$. Each transformer encoder layer contains a self-attention module with $12$ heads. The sizes of the feed-forward layers in each Transformer block are $3072$. The model is optimized with the $L1$ loss using the Adam optimizer with a learning rate of $1e^{-4}$ and a batch size of $32$ for $60$ epochs. Before training, we split the Privacy-risk Saliency Map dataset into a training and validation sets of approximately $45,000$ and $5,000$ samples, respectively. We perform model selection based on the lowest loss on the validation set to avoid over-fitting. For downstream tasks, we follow the training configurations provided in \cite{yang2021superb}. 

\begin{figure*}[t]
\centering
\includegraphics[width=\linewidth]{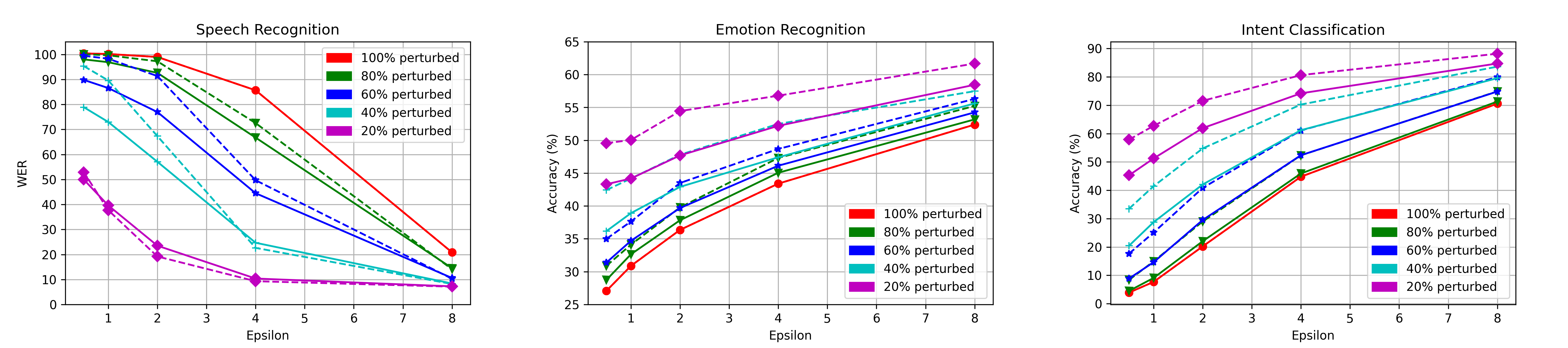}
\vspace{-8mm}
\caption{Proposed framework utility performance. Dashed lines represent the ``random perturbed positions" baselines.}
\label{fig:utility}

\end{figure*}

\begin{figure}[htp]

\centering
\includegraphics[width=0.8\linewidth]{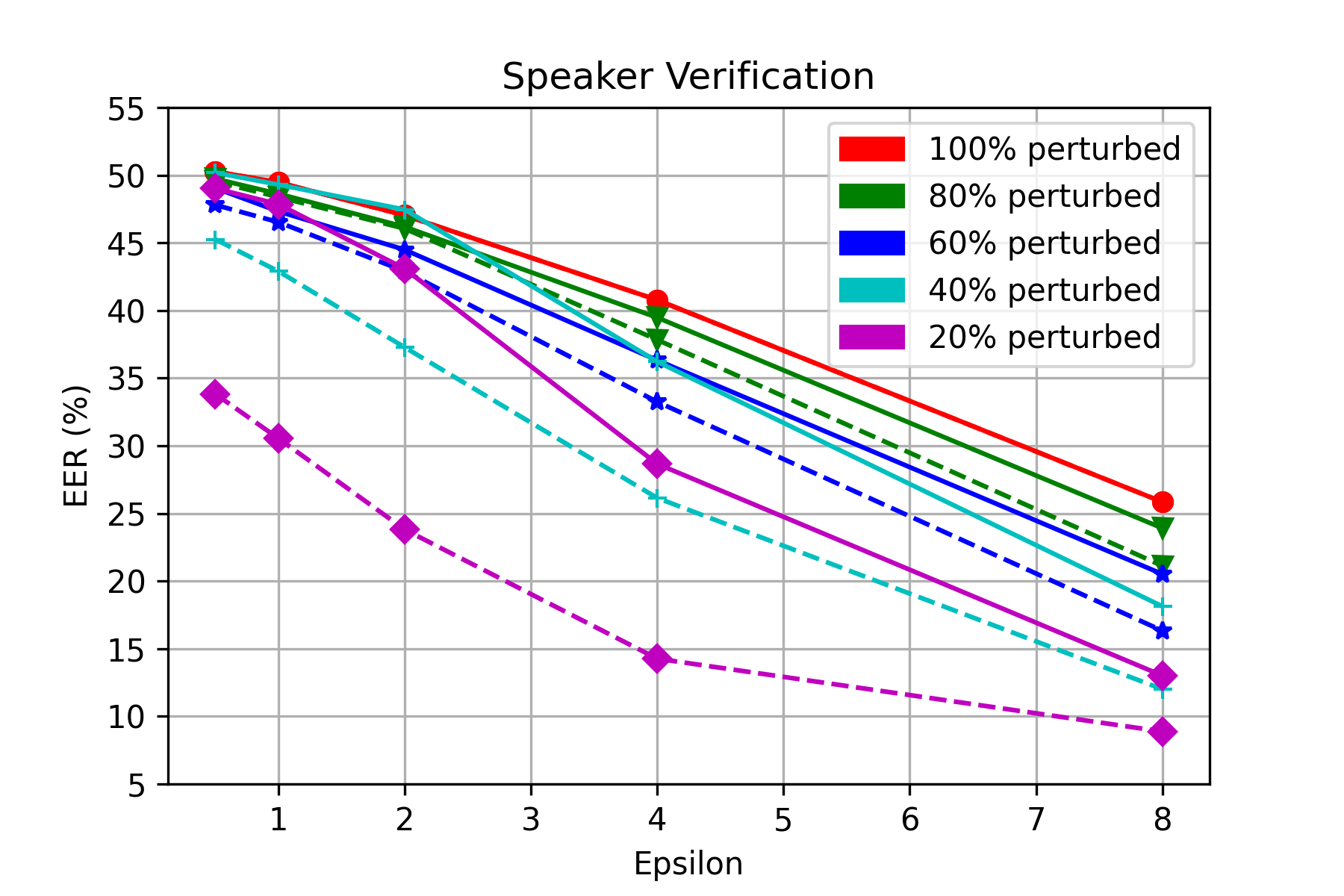}
\vspace{-4mm}
\caption{Proposed framework privacy performance. Dashed lines represent the ``random perturbed positions" baselines.}
\label{fig:privacy}

\end{figure}
\noindent \textbf{Evaluation.} For each task, we train a single model on the extracted features of \texttt{wav2vec2} on the original utterances from the corresponding dataset, and use the trained model for assessing the privacy-utility performances of both our framework and the baselines. We report the performance of these models in the \textit{Original} row in Table \ref{tab:baselines}. 
Assuming an \textit{Ignorant Attack} model might underestimate the risk of speaker-re-identification, however, we are unable to train separate ASV models given the large number of $(k,\epsilon)$ combinations (25 in total) and computation constraints. Training a single ASV (or ASR) model takes around 2 days on a Tesla V100 GPU. Moreover, although the model is unaware of the added noises to the representation, it is also not aware that the input speech are synthesized, making the comparisons fairer. We believe that the EER evaluated by the \textit{Ignorant Attack} model can still inform crucial information about the relative privacy rankings of different methods. For example, as shown in Table \ref{tab:baselines}, \textit{Baseline 1} is more privacy-preserving than \textit{Baseline 2}, which is consistent with the findings in \cite{tomashenko2022voiceprivacy}. It is also important to note that the Lazy-informed and Semi-informed attack models provided in the VoicePrivacy2022 Challenge are not applicable to our work as we are evaluating ASV at a feature-level, \textit{i.e.}, we do not produce speech outputs. 

\section{Results and Discussion}
\begin{table}[t]
\centering 
\begin{tabular}{ccccc}
\hline
           & ASV   & IC & ER & ASR  \\ \hline
Original   & 6.30   & 92.67    & 64.81    & 6.54     \\ \hline
Baseline 1 & 31.63 & 86.97    & 37.73    & 10.70     \\ 
Baseline 2 & 27.18 & 57.84    & 43.89    & 23.65    \\ 
Baseline 3 & 21.60  & 82.78    & 48.49    & 8.52     \\ 
Baseline 4 & 45.40  & 20.01    & 31.18    & 77.12    \\ \hline
Ours ($k=20\%,\epsilon=4$) & 28.66  & 74.19    & 52.20    &  10.38     \\ 
Ours ($k=20\%,\epsilon=1$) & 47.85  & 51.23    & 44.18    & 39.60   \\ \hline
\end{tabular} 
\caption{Performance comparisons between the proposed method and the baselines.}
\label{tab:baselines}
\end{table}

Table \ref{tab:baselines} shows the privacy-utility evaluation for the baseline models. Figure \ref{fig:utility} and \ref{fig:privacy} show the evaluation results for the utility and privacy metrics, respectively. To highlight our PSE's ability to correctly identify positions with \textit{sensitive} information, we add another baseline that randomly chooses $k\%$ of the positions in the extracted \texttt{wav2vec2} representations for perturbation with similar settings to our approach.

We can observe that the proposed approach enables a very flexible range of privacy-utility trade-offs, with EER ranging from $0.13$ to $0.50$ while the utility metrics ranging from random guessing performance to approximately the original performance. On the higher EER range ($>40\%$), the approach ($k=20\%,\epsilon=1$) shows better utility performance (51.23\%, 44.18\%, and 39.60\% in IC, ER, and ASR respectively) compared to \textit{Baseline 4}. Compared to \textit{Baseline 1} and \textit{Baseline 2} in the lower EER range (around 30\%), our approach achieves superior ER performance (52.20\% vs. 43.89\%), comparable ASR WER performance (10.38\% vs. 10.7\%), and lower IC performance (74.19\% vs. 86.97\%) with $k=20\%,\epsilon=4$. In general, feature-level anonymization can be a competitive alternative to voice anonymization when voice outputs are not required, \textit{e.g.,} supervised machine learning applications.

From Figure \ref{fig:privacy}, we can see that our method achieves superior (higher) EER curves compared to the random perturbation baselines in all $\epsilon$ and perturbation ratio variations. It is also important to note that the gap between our method and the random perturbation baselines gets smaller as the proportion of perturbed positions $k$ gets larger, with the most significant gap observed at $k=20\%$. These confirm our assumption that certain positions within the extracted representations carry more sensitive information, and show that our PSE is able to identify high privacy-risk positions for perturbations. However, from Figure \ref{fig:utility}, we can see that the dashed curves generally have better utility performances compared to the solid curves, with the exception of Speech Recognition with a reversed effect. This implies that the selected positions from PSE might potentially be also high-utility positions for downstream tasks.

\section{Conclusion}
In this study, we propose a framework to perturb the extracted representations from pre-trained speech models, with the goal of generating privacy-preserving representation for data sharing. The framework consists of three stages: generating a privacy-risk saliency map dataset for representations extracted from a selected pre-trained model, training a Privacy-risk Saliency Estimator (PSE), and perturbing the extracted representations based on the selected positions from PSE. We validate our framework on a wide range of downstream tasks and find that sharing speech data in the form of extracted representations is competitive to the existing approaches that share speech via voice conversion or voice modification. 
We hope the results motivate further exploration in the direction of privacy-preserving pre-trained representation learning.


\bibliographystyle{IEEEbib}
\bibliography{refs}

\end{document}